\documentstyle[12pt]{article}
\begin{document}
\newcommand{\bea}{\begin{eqnarray}}
\newcommand{\eea}{\end{eqnarray}}
\newcommand{\be}{\begin{equation}}
\newcommand{\ee}{\end{equation}}
\newcommand{\non}{\nonumber}
\global\parskip 6pt
\begin{titlepage}
\vskip .5in
\begin{center}
{\Large\bf Combinatorial Invariants from Four }\\
\vskip .25in
{\Large\bf Dimensional Lattice Models}\\
\vskip 1in
Danny Birmingham \footnote{Supported by Stichting voor Fundamenteel
Onderzoek der Materie (FOM)\\
Email: Dannyb@phys.uva.nl}     \\
\vskip .10in
{\em Universiteit van Amsterdam, Instituut voor Theoretische Fysica,\\
1018 XE Amsterdam, The Netherlands} \\
\vskip .50in

Mark Rakowski\footnote{Email: Rakowski@yalph2.bitnet}   \\
\vskip .10in
{\em Yale University, Center for Theoretical Physics,\\ New Haven,
CT 06511, USA}  \\
\end{center}
\vskip .10in
\begin{abstract}
We study the subdivision properties of certain lattice
gauge theories based on the groups $Z_{2}$ and $Z_{3}$, in four
dimensions. The Boltzmann weights are shown to be invariant under all
type $(k,l)$ subdivision moves, at certain discrete values of the
coupling parameter. The partition function then provides a
combinatorial invariant of the underlying simplicial complex, at
least when there is no boundary. We also show how an extra phase
factor arises when comparing Boltzmann weights under the Alexander
moves, where the boundary undergoes subdivision.
\end{abstract}
\vskip .5in
\begin{center}
ITFA-93-06\,/\, YCTP-P6-93 \\
March 1993
\end{center}
\end{titlepage}

\section{Introduction}

     The application of quantum field theory to the study of
certain problems in topology has been a very fruitful one; we refer the
reader to \cite{BBRT} for a general review of this subject. To state
things quite simply, it has been possible to compute a variety of
topological invariants as correlation functions in special quantum
field theories. While many of these applications employ continuum field
theory techniques, the lack of a precise formulation of quantum
field theory is a serious handicap when ones goal is ultimately to
provide self-contained rigorous argument. These difficulties might be
sidestepped if
discrete lattice models can be employed in ways which avoid the
continuum limit.

     In \cite{top1}, we presented a class of lattice gauge theories
which enjoyed some novel properties under lattice subdivision. The models
were defined on a triangulated 4-manifold with boundary, in terms of
compact Wilson variables. While we
were motivated by some discrete structures which had a formal resemblance
to the pure Chern-Simons theory \cite{AS,EWit}, once the model is
defined no further reference to any continuum theory need be made. We
found that it was possible to define models based on the gauge groups
$Z_{2}$ and $Z_{3}$ in which the Boltzmann weight was invariant under
type 4 Alexander subdivision, at certain discrete values of the coupling
parameter. These observations came as a result of computer studies,
and we were able to present some exact calculations using Mathematica
\cite{Math}.

     Here, we will present analytic proofs of the subdivision properties
of the models defined in \cite{top1}. We will show that the Boltzmann
weight of these theories is invariant under all type $(k,l)$ subdivisions
\cite{gross} of the underlying simplicial complex. This provides
one with a partition
function which is a combinatorial invariant of that complex, at least
in the absence of a boundary. We also show how the Boltzmann weights
behave under Alexander subdivision \cite{Alex}, where the
boundary itself is
subdivided. We find that there is a phase factor associated with these
general subdivision moves at the level of the Boltzmann weights.

     Other topological lattice models have been  formulated previously
\cite{TV,HO,crane}, and we should say a word about them here. While these
theories are also formulated in terms of a triangulation, and an analysis
of subdivision properties has been given, we are not aware of any
connection with the models considered in this paper.
In particular, the Boltzmann weights of these other models are
assembled from the $6j$ symbols, and their extensions.
In \cite{DW,Alt}, certain
Chern-Simons type theories were constructed for
finite groups. It is unclear whether a relation exists between those
models and our four dimensional theory defined  on a manifold with boundary.

    The next section begins with a review of simplicial complexes and
lattice gauge theory, followed by a definition of the models we consider.
An overview of subdivision moves is also given. We then present our main
results regarding $(k,l)$ subdivision invariance, and their proof.
Alexander subdivision is then examined in these models, and we
remark on some issues that arise with different groups. We close
then with a few comments.

\section{Definition of the Models}

To begin, let us recall the basic elements of the theory of
simplicial complexes; we refer to
\cite{JM}, which will be the source for our definitions and notation,
for further details.

Let $\{ v_{0},\cdots ,v_{n}\}$ be a geometrically independent set of points
in some ambient euclidean space $R^{N}$. An  n-simplex spanned by this set
of vertices, is the set of points $x$ of $R^{N}$ which satisfies,
\bea
x &=& \sum^{n}_{i=0}\; t_{i}\, v_{i}\;\; , \;\; with\;\; \\
1 &=& \sum^{n}_{i=0}\; t_{i}\;\; , \;\; and \;\;t_{i}
\geq 0\;\; for\;\; all\;\;
i\;\;.\nonumber
\eea
Pictorially, these can be regarded as points, line segments, triangles, and
tetrahedrons for $n$ equals zero through three respectively.
A simplex which is spanned by any subset of the vertices is called a face
of the original simplex.
An  orientation of a simplex is a choice of ordering of its vertices,
and we let
\bea
[v_{0},\cdots ,v_{n}]
\eea
denote the oriented simplex with the orientation class given by the ordering
$v_{0} \cdots v_{n}$.

A simplicial complex $K$ in $R^{N}$ is a
collection of simplices which are glued together under two restrictions.
Any face of a simplex in $K$ is also in $K$, and the intersection
of any two simplices in $K$ must be  a face of each of them.
The picture here is that
of a collection of simplices glued together under the above restriction.
We will think of a spacetime manifold as being approximated by a certain
simplicial complex.

One defines the boundary operator $\partial_{p}$
on $\sigma=[v_{0},\cdots,v_{p}]$ by:
\bea
\partial_{p}\,\sigma = \sum^{p}_{i=0}\; (-1)^{i}\, [v_{0},\cdots,
\hat{v}_{i},\cdots,v_{p}]\;\;,
\eea
where the `hat' indicates a vertex which has been omitted. It
is easy to show that the composition of boundary operators is zero.

In a Wilson formulation of lattice gauge theory, the basic
dynamical variables
are given  by group valued maps on the 1-simplices (denoted $[a,b]$)
with the rule
that $U_{ba} = U^{-1}_{ab}$. A configuration of the system is then
specified by a collection of group elements, one for each ``link''.
One has, in addition, a gauge group associated to each of the vertices,
and the action of that group on the link variables is defined by,
\be
U_{ab} \rightarrow g_{a}\; U_{ab}\; g^{-1}_{b}\;\;,
\ee
where $g_{a}$ is a group element associated with the vertex $a$.
This group action is also called a gauge transformation.

Given a compact gauge group $G$, together with an invariant measure,
one can define a theory with partition function
\bea
Z = \prod_{\alpha} \int \; dU_{\alpha}
\;\exp [\beta\, S(U) ]\;\;,\label{part}
\eea
where the action functional $S$ of the theory is taken to be a gauge
invariant function
of the link variables defined above,
and the index $\alpha$ indicates the set of independent $1$-simplices
in the simplicial complex $K$.
In the case of a discrete gauge group, the group integration (whose volume
we normalize to unity) is a discrete sum,
\bea
\int \, dU \rightarrow \frac{1}{|G|}\, \sum_{U}\;\; ,
\eea
where $|G|$ denotes the order of the group.
One can also define correlation functions of the link variables,
\bea
<U_{\gamma_{1}}\cdots U_{\gamma_{p}} > = \prod_{\alpha} \int\; dU_{\alpha}\;
U_{\gamma_{1}}\cdots U_{\gamma_{p}}\; \exp [ \beta\, S(U) ]\;\;.
\eea
It should be emphasized that, in general, all of these quantities
depend not only on the
coupling parameter $\beta$, but also on the simplicial complex $K$.

A central role in the construction of lattice gauge theory actions
is played by the holonomy.
Let
$U_{abc} = U_{ab}\, U_{bc} \, U_{ca}$ be the holonomy
based at the first
vertex $a$, around  the triangle determined by $a$, $b$ and $c$, and
traversed in the order from left to right .

We take the action of our theory to be given by,
\be
S=\sum \;\; (U-U^{-1})\star (U-U^{-1})\;\;,\label{a1}
\ee
where $U$ is the above holonomy combination, and
the sum here is over all elementary $4$-simplices in
the simplicial complex. A matrix trace is also to be included for
the case of non-Abelian Lie groups.
The $\star$-product \cite{BR}, to be
recalled presently, is designed to capture some of the properties of
the wedge product of differential forms. In a continuum limit, the
quantity $U-U^{-1}$ becomes proportional to the curvature of a connection,
and (\ref{a1}) then goes over to the Chern form. We mention this from
a purely motivational standpoint; we will not make any use of
continuum theories in this paper. Let us now recall the
definition of the star product \cite{BR}.

The star product is a variant of the usual cup product of cochains
which achieves graded commutativity at the expense of associativity.
Let $c^{r}$ and $c^{s}$ be two maps from the set of oriented $r$- and
$s$-simplices respectively, into a group, and let $< c^{r},[v_{0},
\cdots,v_{r}]>$ represent the evaluation of this map on the particular
r-simplex $[v_{0},\cdots,v_{r}]$. In our applications, we have been
using the notation $U_{abc}$ for the quantity $< U, [abc] >$.
Denote
by $P$, one of the $(r+s+1)!$ permutations of the set of vertices $\{
v_{0},\cdots,v_{r+s}\}$, which span some $(r+s)$-simplex,
and by $Pv_{i}$ the value of that permutation on
$v_{i}$. The star product of $c^{r}$ and $c^{s}$ is defined by,
\bea
&\phantom{.}& < c^{r}\star c^{s}, [v_{0},\cdots,v_{r+s}] > \; =\\
&\phantom{.}& \frac{1}{(r+s+1)!}\,
\sum_{P}\; (-1)^{| P |}\; < c^{r},[Pv_{0},\cdots,Pv_{r}] >\;
\cdot < c^{s},[Pv_{r},\cdots, Pv_{r+s}] >\;\;,\nonumber
\eea
when the order $v_{0}\cdots v_{r+s}$ is in the equivalence class of the
orientation of the simplex $[v_{0},\cdots,v_{r+s}]$ (this determines the
overall sign of the product), and
where the sum is over all permutations of the vertices.
The actual number
of independent terms in this sum is given  by the number of ways one can
partition the set of vertices into two parts which contain one vertex in
common, and an easy counting yields
\bea
\frac{(r+s+1)!}{r!\, s!}\;\;.
\eea

As we have seen, the holonomy $U$ is a group valued map on $2$-simplices,
and therefore the above action is naturally defined on
a $4$-simplex.
One should note that the quantity $(U-U^{-1})_{abc}$ enjoys the property
of antisymmetry in its last two indices; this is a simple consequence of
the relation $U_{abc} = U^{-1}_{abc}$. When the group is Abelian, one
has, moreover, antisymmetry in all three indices.
The star product in our action, when evaluated on a given 4-simplex,
leads generically to 5! terms, however, the
symmetries present in (\ref{a1}) reduce that number to 15 distinct
combinations.
The Boltzmann weight for this theory, evaluated on
the simplex $[v_{0},v_{1},v_{2},v_{3},v_{4}]$,
can now be written in the form:
\bea
W[v_{0},v_{1},v_{2},v_{3},v_{4}]&
=& B[v_{0},v_{1},v_{2},v_{3},v_{4}]\,
B[v_{0},v_{1},v_{3},v_{4},v_{2}]\,B[v_{0},v_{1},v_{4},v_{2},v_{3}]\non\\
& &B[v_{1},v_{0},v_{2},v_{4},v_{3}]\,
B[v_{1},v_{0},v_{3},v_{2},v_{4}]\,B[v_{1},v_{0},v_{4},v_{3},v_{2}]\non\\
& &B[v_{2},v_{0},v_{1},v_{3},v_{4}]\,
B[v_{2},v_{0},v_{3},v_{4},v_{1}]\,B[v_{2},v_{0},v_{4},v_{1},v_{3}]\non\\
& &B[v_{3},v_{0},v_{1},v_{4},v_{2}]\,
B[v_{3},v_{0},v_{2},v_{1},v_{4}]\,B[v_{3},v_{0},v_{4},v_{2},v_{1}]\non\\
& &B[v_{4},v_{0},v_{1},v_{2},v_{3}]\,
B[v_{4},v_{0},v_{2},v_{3},v_{1}]\,B[v_{4},v_{0},v_{3},v_{1},v_{2}]
\;\;,\non\\
\label{bw}
\eea
where,
\bea
B[v_{0},v_{1},v_{2},v_{3},v_{4}] =
\exp [\beta\, (U-U^{-1})_{v_{0}v_{1}v_{2}}\,
(U-U^{-1})_{v_{0}v_{3}v_{4}} ]\;\; .
\eea
One general feature worth observing is that each $B$ factor depends
on two independent holonomies which share a common vertex; the term
``bowtie'' seems appropriate to describe this structure.
In the following, our analysis shall proceed for general complex
coupling $\beta$.

Our main concern here is to study these models for the case of
the discrete Abelian groups $Z_{p}$. However, an action which
depends on the combination $(U-U^{-1})$ necessarily leads to a
trivial theory for the case of $Z_{2}$.
One may then wish to consider the action
\be
S =\sum \;\; (U-1)\star (U-1)\;\;.
\ee
However, for Abelian groups, the holonomy $U_{abc}$ is invariant under
cyclic permutations of the indices. In addition, for the case of $Z_{2}$,
we have the relation $U = U^{-1}$, for all group elements. As a result,
the holonomy combination is in fact symmetric in all indices, and the
action above vanishes. Nevertheless, as shown in \cite{top1}, one can
simply define the Boltzmann weight for a given $4$-simplex to be of
the form (\ref{bw}), with
\be
B[v_{0},v_{1},v_{2},v_{3},v_{4}] = \exp[\beta(U-1)_{v_{0}v_{1}v_{2}}
(U-1)_{v_{0}v_{3}v_{4}}]\;\;.
\ee

One can proceed and compute the partition function
for these theories defined on  a simplicial complex $K$.
We wish to study, however, the behavior of this function under
subdivision of
the complex. Let us now recall two bases of subdivision operations which
accommodate an analysis of this question.

{\em The Alexander Moves:}\\
Consider a single $4$-simplex $[v_{0},v_{1},v_{2},v_{3},v_{4}]$.
The subdivision operations of Alexander type can in turn be described
as follows.

Type 1 Alexander subdivision:
\begin{eqnarray}
[ v_{0},v_{1},v_{2},v_{3},v_{4} ] \rightarrow [ x,v_{1},v_{2},v_{3},
v_{4} ] + [v_{0},x,v_{2},v_{3},v_{4} ]\;\;,
\end{eqnarray}

Type 2 Alexander subdivision:
\begin{eqnarray}
[ v_{0},v_{1},v_{2},v_{3},v_{4}] \rightarrow
[ x,v_{1},v_{2},v_{3},v_{4} ] + [ v_{0},x,v_{2},v_{3},
v_{4} ] + [v_{0},v_{1},x,v_{3},v_{4} ]\;\;,
\label{as2}
\end{eqnarray}

Type 3 Alexander subdivision:
\begin{eqnarray}
[ v_{0},v_{1},v_{2},v_{3},v_{4}] &\rightarrow&
[ x,v_{1},v_{2},v_{3},v_{4} ] + [ v_{0},x,v_{2},v_{3},
v_{4} ] + [v_{0},v_{1},x,v_{3},v_{4} ]\non\\
&+& [ v_{0},v_{1},v_{2},x,v_{4} ]\;\;,
\end{eqnarray}

Type 4 Alexander subdivision:
\begin{eqnarray}
[ v_{0},v_{1},v_{2},v_{3},v_{4}] &\rightarrow&
[ x,v_{1},v_{2},v_{3},v_{4} ] + [ v_{0},x,v_{2},v_{3},
v_{4} ] + [v_{0},v_{1},x,v_{3},v_{4} ] \non\\
&+& [ v_{0},v_{1},v_{2},x,v_{4} ] + [ v_{0},v_{1},v_{2},v_{3},x ]\;\;.
\end{eqnarray}

One can picture the move of type $1$ as the introduction of an
additional vertex
$x$, which is placed at the center of the $1$-simplex
$[v_{0},v_{1}]$, and is then joined to all the remaining vertices
of the $4$-simplex.
Moves $2$ to $4$ involve a similar construction, where $x$ is placed
at the center of the simplices $[v_{0},v_{1},v_{2}]$, $[v_{0},v_{1},v_{2},
v_{3}]$, and finally $[v_{0},v_{1},v_{2},v_{3},v_{4}]$.
There is, in addition, a type $0$ move which is effected by replacing
a vertex of the simplicial complex by a new vertex. This can be
considered as  a degenerate case, and need not concern us in the following.

According to Alexander \cite{Alex},
two simplicial complexes are said to be equivalent if and only if it is
possible to transform one into the other by a sequence of these moves.
Hence, any function of $K$ which is invariant under these moves
yields a combinatorial invariant of the simplicial complex.

{\em The $(k,l)$ Moves:}\\
Another basis of subdivision operations is available,
known as the $(k,l)$ moves, and these
allow for a more convenient analysis.
In particular, it has been shown \cite{gross} that the basis of $(k,l)$ moves
is equivalent to the Alexander moves for the case of
closed manifolds, for dimensions less than or equal to four.
In the four dimensional case under study, we have
five $(k,l)$ moves, with $k=1,\cdots,5$, and $k+l=6$.
It suffices to consider the first three cases; the $(4,2)$ and $(5,1)$
moves are inverse to the $(2,4)$ and $(1,5)$ moves, respectively.

The $(1,5)$ move:
\begin{eqnarray}
[ v_{0},v_{1},v_{2},v_{3},v_{4}] &\rightarrow&
[ x,v_{1},v_{2},v_{3},v_{4} ] + [ v_{0},x,v_{2},v_{3},
v_{4} ] + [v_{0},v_{1},x,v_{3},v_{4} ] \non\\
&+& [ v_{0},v_{1},v_{2},x,v_{4} ] + [ v_{0},v_{1},v_{2},v_{3},x ]\;\;,
\end{eqnarray}

The $(2,4)$ move:
\bea
[v_{0},v_{1},v_{2},x,v_{3}] &+& [v_{0},v_{1},v_{2},v_{3},y] \rightarrow
[v_{0},v_{1},v_{2},x,y]
+ [v_{0},v_{2},v_{3},x,y] \non\\
&+&[v_{0},v_{1},v_{3},y,x]
+[v_{1},v_{2},v_{3},y,x]\;\;,
\eea

The $(3,3)$ move:
\bea
[v_{0},v_{1},v_{2},x,y] &+& [v_{0},v_{1},v_{2},y,z]
+ [v_{0},v_{1},v_{2},z,x]
\rightarrow
[x,y,z,v_{0},v_{1}]  \non\\
&+& [x,y,z,v_{1},v_{2}] + [x,y,z,v_{2},v_{0}]\;\;.
\eea

A simple observation is that the $(1,5)$ move is identical to
the type $4$ Alexander subdivision.
One notes that the simplices on the left hand side of
the $(2,4)$ move share a common $3$-simplex $[v_{0},v_{1},v_{2},v_{3}]$,
while those on the right have a common $1$-simplex $[x,y]$.
For the case of the $(3,3)$ move, the $2$-simplex $[v_{0},v_{1},v_{2}]$
is common to the left hand side,
with $[x,y,z]$ being common to the right.
Furthermore, one can verify that the boundary of the $4$-simplex remains
unchanged as a result of these operations.

{\em Previous Results:}\\
Before proceeding with the general analysis, we recall the results
obtained for the $4$-disk and $4$-sphere, for the groups $Z_{2}$ and
$Z_{3}$.
A complete calculation of the partition function for arbitrary complex
coupling $\beta$ was presented in \cite{top1}.
The central observation was that subdivision invariant points
were present in both models. Indeed, it was explicitly checked that
the partition functions
of the $4$-disk remained invariant under all Alexander moves,
at these special points. It was further shown that the results for
$S^{4}$ were invariant with respect to Alexander type $4$ subdivision.
For each of the models studied, a natural scale factor was present,
and we denote this by
$s(2)= \exp[4\beta]$ and $s(3)=\exp[-3\beta]$
for the case of $Z_{2}$ and $Z_{3}$, respectively.
The corresponding subdivision invariant points are then given when
$s(2)^{2}=1$, and $s(3)^{3}=1$.
A point worth mentioning is that the Boltzmann weights themselves
are group valued at these special points.
Let us quickly summarize some of those results.

Beginning with the $Z_{2}$ theory, we found the partition function
on the $4$-disk to be given by
\be
Z(s(2)) = \frac{1}{2^{4}}(9 + 7s(2))\;\;,
\ee
when $s(2)^{2} =1$. The two roots of unity, $+1$ and $-1$ yield the
values $1$ and $1/8$ respectively.
For the case of the $4$-sphere $S^{4}$, we find that the partition
function assumes a value of $Z=1$, when $s(2)^{2}=1$.

Turning now to the $Z_{3}$ theory, we found the partition function
on the $4$-disk to be:
\be
Z(s(3)) = \frac{1}{3^{4}}(29 + 26 (s(3) + s(3)^{-1}))\;\;,
\ee
when $s(3)^{3} =1$.
The trivial subdivision invariant point $s(3)=1$ yields a value $Z=1$,
while the other two cube roots of unity give a value $Z=1/27$.
Again, for the case of $S^{4}$, one finds a value of $Z=1$ when
$s(3)^{3} =1$.

 These results were obtained through the use of Mathematica \cite{Math}
to evaluate the partition functions. While we could show through
exhaustive computer checks that these models had interesting subdivision
invariant points, a clear analytic
understanding of these special properties was generally lacking.
This we supply in the following sections.

\section{Main Results}

   Having laid the foundational material in the preceding
sections, we can now state and prove the main results.
The aim of this section is to establish the behavior of
the Boltzmann weights under all $(k,l)$ type subdivisions.
In order to treat
both the $Z_{2}$ and $Z_{3}$ models uniformly, it is expedient to let $X$
denote the combinations $U-1$ and $U-U^{-1}$, respectively,
for those models.
It will further be convenient to let $B[0,1,2,3,4]$ represent the expression
$B[v_{0},v_{1},v_{2},v_{3},v_{4}]$; using subscripts to keep track
of vertices should cause no confusion. We begin with a lemma.

{\bf Lemma:} The Boltzmann weights for a given vertex ordering
satisfy the conditions,
\be
B[0,1,2,3,4]\, B[0,1,2,4,5]\,B[0,1,2,5,3] = \exp[\beta\, X_{v_{0}v_{1}
v_{2}} X_{v_{3}v_{4}v_{5}}]
\;\; ,\label{zbbb}
\ee
\be
B[0,1,2,3,4]\,B[1,2,0,4,3] =
\exp[\beta\, X_{v_{0}v_{1}v_{2}}
X_{v_{0}v_{1}v_{4}}] \; \exp[ -\beta \, X_{v_{0}v_{1}v_{2}}
X_{v_{0}v_{1}v_{3}}] \;\;,\label{zbb}
\ee
at the points $s(2)^{2} =1$ and $s(3)^{3}=1$, in the $Z_{2}$ and
$Z_{3}$ theories respectively.

   Consider first the $Z_{2}$ case. One notices immediately that the
relation (\ref{zbbb}) is trivially satisfied for
$U_{v_{0}v_{1}v_{2}} = 1$, so
we only need to consider the case $U_{v_{0}v_{1}v_{2}} = -1$. For
simplicity of notation, let $x= U_{v_{0}v_{3}v_{4}}$, $y= U_{v_{0}v_{4}
v_{5}}$, and $z = U_{v_{0}v_{5}v_{3}}$. Noticing that $U_{v_{3}
v_{4}v_{5}} = x\, y \, z$, our assertion is then equivalent to,
\bea
1 = \exp[-2\, \beta \,( (x-1) + (y-1) + (z-1) - (xyz - 1))]\;\; .
\label{xyz2}
\eea
Now, $x$, $y$, and $z$ are independent group elements, and the image
set of the function
\bea
(x,y,z) \rightarrow (x-1) + (y-1) + (z-1) - (xyz - 1)
\eea
is easily seen to be $\{ 0, -4 \}$. Recalling that $s(2) = \exp[4\,\beta]$,
one sees then that (\ref{xyz2}) is satisfied at $s(2)^{2}=1$.

   The $Z_{3}$ relation follows in the same way; here we need only check the
case $U_{v_{0}v_{1}v_{2}} = \exp[ \pm 2\pi i/3]$. The assertion
(\ref{zbbb}) is then equivalent to,
\bea
1 = \exp[ \pm \beta i \sqrt{3}\, ( (x-x^{-1}) + (y-y^{-1}) +
(z-z^{-1}) - (xyz - x^{-1}y^{-1}z^{-1})) ]\;\; ,\label{xyz3}
\eea
where $x=U_{v_{0}v_{3}v_{4}}$, $y=U_{v_{0}v_{4}v_{5}}$, and $z=
U_{v_{0}v_{5}v_{3}}$ are independent $Z_{3}$ elements. The image set
of the function,
\bea
(x,y,z)\rightarrow (x-x^{-1}) + (y-y^{-1}) + (z-z^{-1}) -
(xyz - x^{-1}y^{-1}z^{-1})\;\; ,
\eea
is easily seen to be $\{ 0, \pm 3i\sqrt{3} \}$. With $s(3) = \exp[
-3 \beta]$, one then finds that (\ref{xyz3}) is satisfied at the points
$s(3)^{3} =1$.

    The proof of the second relation (\ref{zbb}) is very similar, and
we omit the details. Our main result is the following theorem.

{\bf Theorem:} The full Boltzmann weights satisfy the relation,
\bea
W[0,1,2,3,4]\,W[0,1,2,4,5]\,W[0,1,2,5,3]
&=&W[0,1,3,4,5]\,W[1,2,3,4,5]\non\\
&\,&W[2,0,3,4,5] \;\;,\label{WWW}
\eea
at the points $s(2)^{2}=1$ and $s(3)^{3}=1$, in the $Z_{2}$ and
$Z_{3}$ theories respectively.

The proof of this, while straightforward, is surprisingly tedious. Each
of the $W$ factors is itself a product of 15 factors. One can write
out all 90 $B$ factors that occur in (\ref{WWW}), and methodically
use the the identities established in the lemma to verify the claim.
In our analysis, we used the identity,
\bea
B[0,1,2,3,4] \, B[0,1,2,4,5]\, B[0,1,2,5,3]
&=&B[3,4,5,0,1] \, B[3,4,5,1,2]\non\\
&\,& B[3,4,5,2,0]\;\;,
\eea
which is a trivial consequence of (\ref{zbbb}), to eliminate all but
18 of the 90 terms. The identity (\ref{zbb}) was then used to polish
off the remaining factors.

   Armed with this theorem, is is now a simple matter to understand the
subdivision properties of the Boltzmann weights under the remaining
moves.

{\bf Corollary:} The full Boltzmann weights satisfy the following two
relations:
\bea
W[0,1,2,3,4]\,W[0,1,2,5,3] &=& W[1,2,3,4,5]\, W[2,0,3,4,5]\,
W[0,1,3,4,5]\,\non\\
&\,&W[1,0,2,4,5]\;\;,
\eea
\bea
W[0,1,2,3,4]&=&W[5,1,2,3,4]\, W[0,5,2,3,4]\, W[0,1,5,3,4]
\,W[0,1,2,5,4]\non\\
&\,& W[0,1,2,3,5] \;\; ,
\eea
at the points $s(2)^{2} = 1$ and $s(3)^{3}=1$ for the groups $Z_{2}$
and $Z_{3}$ respectively.

   The proof here is a simple application of the theorem, together
with the fact that,
\bea
W[0,1,2,3,4]^{-1} = W[0,1,2,4,3]
\eea
in our theories; this relation holds at $s(2)^{2} =1$ in the $Z_{2}$ case,
and quite generally in the $Z_{3}$ model. This means that $W$ is actually
invariant under a reversal of orientation at the special points in the
$Z_{2}$ model, and is exchanged for its inverse in all models based on
the action (\ref{a1}).

As a consequence of these results, one can immediately establish
the fact that the partition function for these models provides
a combinatorial invariant of an arbitrary simplicial complex, at
least for the case of zero boundary. In particular, we can now assert
that the results presented previously \cite{top1} for the case of
$S^{4}$ do indeed correspond to a combinatorial invariant.

\section{Behavior under the Alexander Moves}

The subdivision moves introduced by Alexander, and reviewed in an earlier
section, are slightly more complicated. These moves generally induce
subdivisions of the boundary of a given 4-simplex. Nevertheless, our
understanding of the type $(k,l)$ subdivision will allow us to fully analyze
these other moves.

Consider the type 3 Alexander move where we add the $v_{5}$ vertex
to the center of $[v_{0},v_{1},v_{2},v_{3}]$. Using subscripts once again to
keep track of the vertices, this move takes the form:
\bea
[0,1,2,3,4] \rightarrow [5,1,2,3,4] + [0,5,2,3,4] + [0,1,5,3,4] +
[0,1,2,5,4] \;\; .\label{A3}
\eea
It is useful to note how the boundary transforms under this move;
a simple check reveals that the boundary component $[0,1,2,3]$
undergoes a three dimensional Alexander type $3$ subdivision, namely,
\be
[0,1,2,3] \rightarrow [5,1,2,3] + [0,5,2,3] + [0,1,5,3] + [0,1,2,5]
\;\;.
\ee
The fundamental question is how $W[0,1,2,3,4]$ is related to the weights
of the four 4-simplices on the right hand side of equation (\ref{A3}).
Again,
the $(3,3)$ identity we established in the last section proves to be
the key to resolving this. It is a quick exercise to show that,
\bea
W[0,1,2,3,4] &=& W[5,0,1,2,3]\, ( \, W[5,1,2,3,4]\,  W[0,5,2,3,4] \,
                                   W[0,1,5,3,4]\, \non\\
& &W[0,1,2,5,4] \,)\;\; .
\eea
The Boltzmann weight $W$ is not invariant under this move, but it picks
up what one might wish to view as a phase factor associated with adding
the $v_{5}$ vertex to the center of $[v_{0},v_{1},v_{2},v_{3}]$; the
``phase" being the quantity $W[5,0,1,2,3]$.
It is equally simple to
write the corresponding ``phases" associated with the type 1 and 2 Alexander
moves, though we won't catalogue them here. The type 4 move is identical
to $(1,5)$, and we know that there is no phase factor in that case.

\section{$Z_{4}$ and Beyond}

One rather immediate question about the results we have laid out in the
previous two sections would be with regard to extending them to the
general $Z_{p}$ case. This is not automatic, and an analysis of the group
$Z_{4}$ already begins to show a departure from what happened for $Z_{2}$
and $Z_{3}$. If one repeats the same analysis for a $Z_{4}$ type theory
defined by the action (\ref{a1}), one finds that not all fourth roots
of unity yield subdivision invariant points under the $(k,l)$ moves.
Defining the analogous scale factor to be $s(4) = \exp[ -4\beta ]$, one
finds only the two points corresponding to $s(4)^{2}=1$.
A similar situation arises for $Z_{6}$; with the scale factor denoted by
$s(6) =\exp[-3\beta]$, one finds subdivision invariant points when
$s(6)^{3}=1$.
For $Z_{5}$, however,
the entire structure of the theory is rather more complicated, and it is
an open question as to whether one can find subdivision invariant points.
Equally, we must leave extensions to other types of groups, both
discrete and continuous, for future investigation.

\section{Concluding Remarks}
Having established the combinatorial invariance of the partition
function for the $Z_{2}$ and $Z_{3}$ models, perhaps the most pressing
issue is to determine the precise nature of
this invariant. In particular, it is of interest to explicitly
compute the invariant for a variety of closed manifolds.
As we have seen, the Boltzmann weight is invariant
under the Alexander moves, up to certain ``phase" factors associated
with the subdivision induced on the boundary. Our conclusion thus
falls short of declaring the partition function to be a combinatorial
invariant for a four dimensional manifold with boundary.
Nevertheless, as we have seen from our computer studies, the partition
function on the $4$-disk is indeed invariant under
all subdivision moves of Alexander type. Further work is required
in this arena.
One might also hope that the natural correlation functions associated
with each of these models may enjoy special properties with
respect to subdivision, but we leave that for the future.

\end{document}